\documentclass{aa}
\input psfig.sty
\begin{document}
\title{Radio Spectrum and Distance of the SNR HB9}
\author{Denis A. Leahy\inst{1}
      \and
        Wenwu Tian\inst{1,2}}
\offprints{W. W. Tian}
\institute{Department of Physics \& Astronomy, University of Calgary, Calgary, Alberta T2N 1N4, Canada\\
\and National Astronomical Observatories, CAS, Beijing 100012, China}

\date{Received xx, 2006; accepted xx, 2006} 

\abstract{New images are presented of the supernova remnant (SNR) HB9 based on  
408 MHz and 1420 MHz continuum emission and HI-line emission data from the Canadian 
Galactic Plane Survey (CGPS).
Two different T-T plot methods and new integrated flux densities give spectral index 
(S$_{\nu}$$\propto$$\nu$$^{-\alpha}$) for the whole of HB9 of 0.48$\pm$0.03; 0.49; and 0.47$\pm$0.06, respectively. 
These values are lower than the previous spectral index estimated for HB9 ($\alpha$=0.61). 
The change is  mainly due to improved 1420 MHz data.  
No difference in spectral index is detected between strong and weak filaments. 
A new result is that the spectral index for interior regions is steeper than for the rim. 
This can be explained by a standard curved interstellar electron energy spectrum combined 
with lower interior magnetic field compared to that near the outer shock. This results in
a larger proportion of steep spectrum emission for lines-of-sight through the central body of the SNR.
HI observations show structures probably associated with the SNR in the 
radial velocity range -3 to -9 km$/$s, suggesting a kinematic distance of 0.8$\pm$0.4 kpc 
for the SNR. This is consistent with the distance to the radio pulsar PSR B0458+46, offset 
from the center of HB9 by 23$^{\prime}$. However the pulsar spindown and kinematic
ages are significantly greater than estimates of the SNR age: the Sedov age for HB9 is 
6600 yr and the evaporative cloud model yields ages of 4000-7,000 yr. 
}
\titlerunning{Radio spectrum and distance of HB9}
\maketitle

\section{Introduction}

The production of high-energy particles in our Galaxy is closely related 
to shock acceleration in supernova remnants (SNR). The study of the radio spectra produced
 by high-energy electrons in SNRs allows one to learn about
the electron energy spectrum. 
Due to its large diameter HB9 is one of the best candidates for a study of SNRs' 
spectral index spatial variations. Variations have been observed to occur in 
some other large angular sized SNRs (Leahy $\&$ Tian 2005; Tian $\&$ Leahy 2005, 
Uyaniker et al. 2004, Alvarez et al. 2001). Spatial variations in the spectral index across HB9 has been previously noted by Leahy et al. (1998) and Leahy $\&$ Roger (1991), but a new set of higher 
resolution and sensitivity observations of the Canadian 
Galactic Plane Survey (CGPS) combined with new spectral 
index analysis methods, impel us to carry out a new spectral index study of HB9.
A lack of direct estimates of the distance to HB9 in 
previous research (Leahy $\&$ Roger 1991, Lozinskaya 1981, Milne 1979) is corrected 
here by an analysis of the new HI observations of HB9.   

\section{Observations and Image Analysis}

The continuum and HI emission data sets come from the CGPS,
described in detail by Taylor et al. (2003).
The data sets are mainly based on observations from the Synthesis Telescope
(ST) of the Dominion Radio Astrophysical Observatory (DRAO). The spatial
resolution of the continuum images of HB9 is 49$^{\prime}$$^{\prime}$ $\times$ 
68$^{\prime}$$^{\prime}$  at 1420 MHz and 2.8$^{\prime}$$\times$3.9$^{\prime}$ at
408 MHz. The synthesized beam for
the HI line images is 58$^{\prime}$$^{\prime}$$\times$ 80$^{\prime}$$^{\prime}$ 
and the radial velocity resolution is 1.32 km$/$s. The continuum images are noise limited with
 rms of $\sim$0.3 
mJy$/$beam at 1420 MHz and $\sim$3 mJy$/$beam at 408 MHz. The DRAO ST observations
are not sensitive to structures larger than an angular
size scale of about 3.3$^{\circ}$ at 408 MHz and 56$^{\prime}$ at 1420 MHz.
Thus the CGPS includes data from the 408 MHz all-sky survey of Haslam et al (1982), 
sensitive to structure greater than 51$^{\prime}$, and the Effelsberg 1.4 GHz Galactic 
plane survey of Reich et al. (1990, 1997), sensitive to structure greater than 
9.4$^{\prime}$. 
The large scale HI data is from the single-antenna survey of the
CGPS area (Higgs $\&$ Tapping 2000) with resolution of 36$^{\prime}$.  

There are many compact sources (CS) overlapping the face of HB9. In order to better 
study the spectral index distribution, we compare two methods 
to reduce the effects of CS on the SNR's spectral index. 
The first removes CS at both frequencies but is limited to the resolution of the 
408 MHz image; the other removes CS only in the 1420 MHz image so is effective
at higher spatial resolution. 
The first method was introduced by Tian and Leahy (2005) and the second was 
introduced by Leahy (2006). Both have been proved effective in removing CS contamination.
We also analyze HI spectral line observations of HB9 and estimate HB9's distance.
 
\section{Results}

\subsection{HB9 Flux Densities at 408 MHz and 1420 MHz}

The first row of Fig. 1 shows the 408 MHz (left) and 1420 MHz (right) images of HB9
overlaid by boxes used for spectral index determination. 
The 408 MHz and 1420 MHz CS-subtracted images are shown in the second row of Fig. 1.
160 CS at 1420 MHz and 61 CS at 408 MHz are detected within HB9. 
The total integrated flux densities of these CS are 12.0 Jy at 408 MHz and 5.6 Jy at 1420 MHz. 
After subtracting the flux of CS, HB9's integrated flux density is 117.8$\pm$5.3 Jy at 
408 MHz and 65.9$\pm$3.4 Jy at 1420 MHz.
This gives an integrated flux-density 408-1420 MHz spectral index $\alpha$=0.48$\pm$0.06 including 
CS, and 0.47$\pm$0.06 after subtracting CS. The results are consistent within errors. 
 
\subsection{HB9 T-T Plot Spectral Indices}

We have applied the 
T-T plot method to investigate spectral indices (e.g. Tian $\&$ Leahy 2005). 
The T-T plot method is first applied to the whole area covering HB9 (the single big box 
of Fig. 1). Fig. 2 gives the 408-1420 MHz T-T plots for three cases (from left to right): 
case 1- maps including CS ($\alpha_{auto}$=0.50$\pm$0.03); case 2- CS subtracted from both 
408 MHz and 1420 MHz maps ($\alpha_{auto}$=0.48$\pm$0.03); case 3- maps with Gaussian 
fits to CS subtracted from the 1420 MHz map only ($\alpha_{manual}$=0.49). 
The detailed description of case 3 is given in Leahy (2006): 
the motivation is two-fold: to remove CS which are only clearly resolved at 1420 MHz 
and not at 408 MHz; and to be able to clearly separate points in the T-T plot which belong 
to CS from points which belong to diffuse SNR emission. The subscript auto refers to the case 
of an automated linear fit
including all of the points, the subscript manual refers to the case of a manual fit done to the points
excluding points due to CS. The points due to CS are clearly seen 
in the left plot of Fig. 2 if they have a significantly different slope than the points from the SNR. 
For the middle plot, CS are subtracted at both 408 and 1420 MHz. However, there are artifacts 
due to some CS subtracted at 1420 MHz but not at 408 MHz and vice versa. 
For the right plot, since CS are subtracted only at 1420
MHz, points due to CS show up as vertical lines of points, with 408 MHz flux but no 1420 MHz flux.
This occurs even if a CS has a spectral index similar to the diffuse SNR emission.
  
To study spatial variations in HB9, we divide it into 52 regions (see the top plots of Fig. 1). 
The case 3 method removes fainter CS detected only at 1420 MHz (for discussion, see Leahy, 2006). 
We give the results in Table 1 and for comparison, the spectral index from case 1 including CS is also given.
Table 1 shows that case 1 and case 3 results can be quite different. 
Fig. 3 shows T-T plots for region 21: for case 1 (left, $\alpha_{auto}=0.51\pm0.01$), case 2 (middle,
$\alpha_{auto}=0.88\pm0.11$), and case 3 (right, $\alpha_{manual}=0.61$). 
The spectral index for case 1 is dominated by the single bright CS.
Case 2 gives an erroneous (too large) value of spectral index.
The reason is that in the presence of diffuse emission, the Gaussian fitting to the strongest 
CS produces very low flux at 408 MHz, resulting in a residual flux in
the CS subtracted map at this frequency.  
The advantage of case 3 is more apparent here, 
since there is no overcrowding of the CS in the T-T plot, like there was in Fig. 2 (right). 
This allows easy identification of the CS, including the faint CS at TB(1420)=5.8K, 

A histogram of the T-T plot spectral index values from Table 1 is shown in Fig. 4. 
The dashed line shows the 
spectral indices obtained from automatic fits, including CS at both 408 MHz and 1420 MHz (case 1); 
the solid line
shows the spectral indices obtained from manual fits for CS subtracted at 1420 MHz (case 3). 
The effect of CS on the spectral index distribution is clear: there is a second peak 
in the histogram introduced at 0.7 to 0.9 due to CS, whereas the SNR only emission (case 3) 
has a smooth distribution of indices between 0.4 and 0.8. 

\begin{figure*}
\vspace{60mm}
\begin{picture}(200,300)
\put(-34,460){\includegraphics{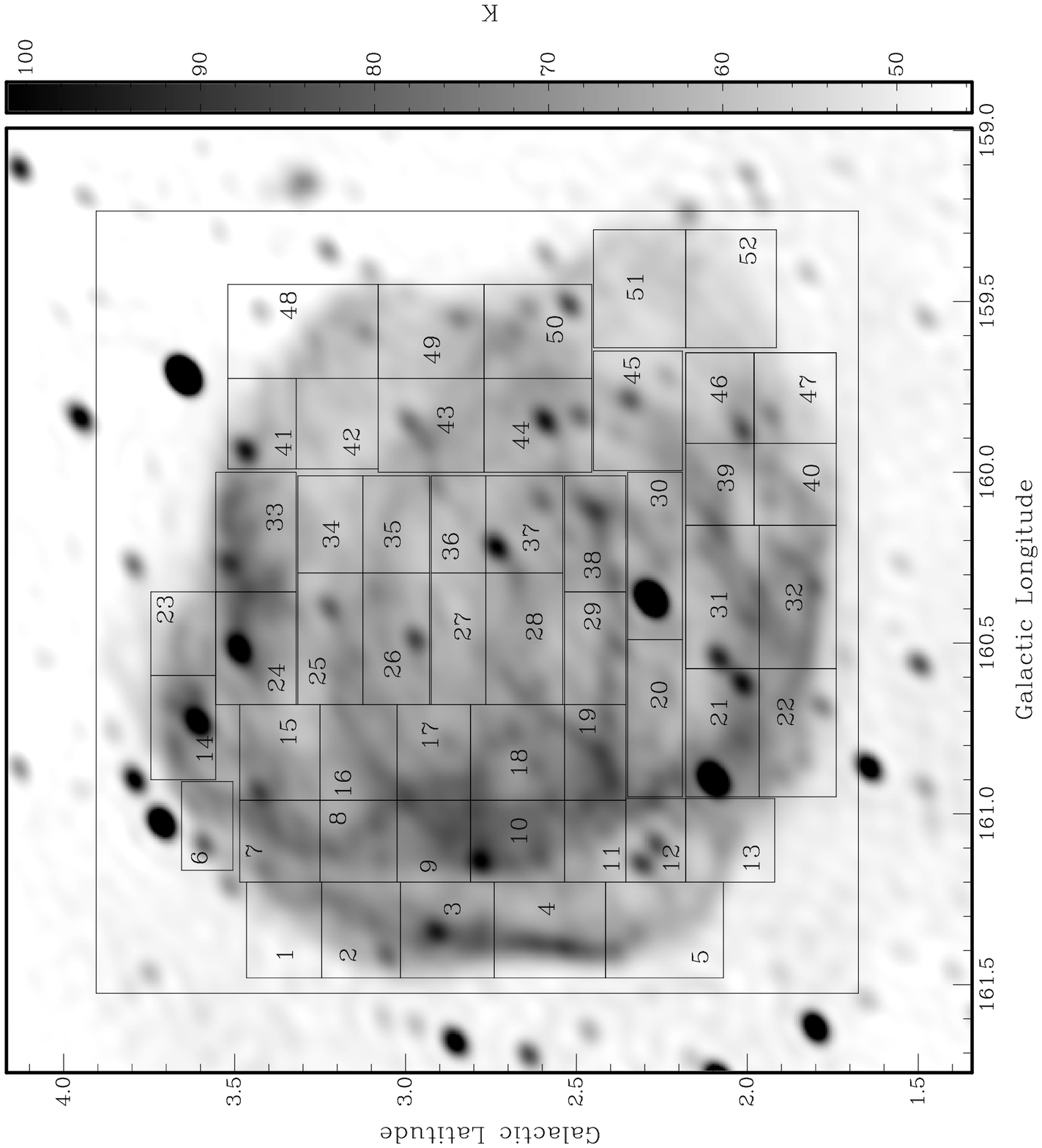}}
\put(230,460){\includegraphics{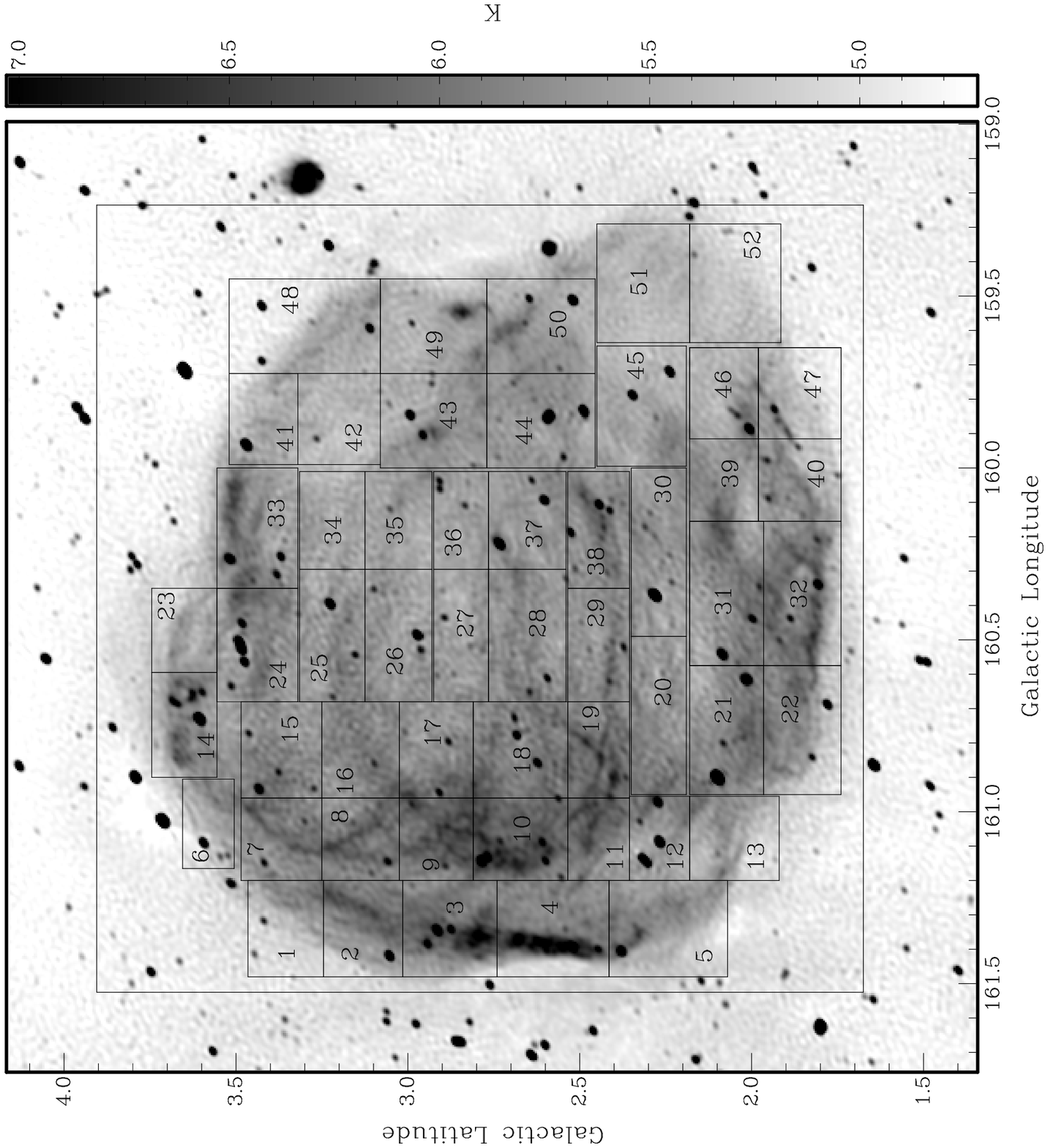}}
\put(-130,300){\includegraphics{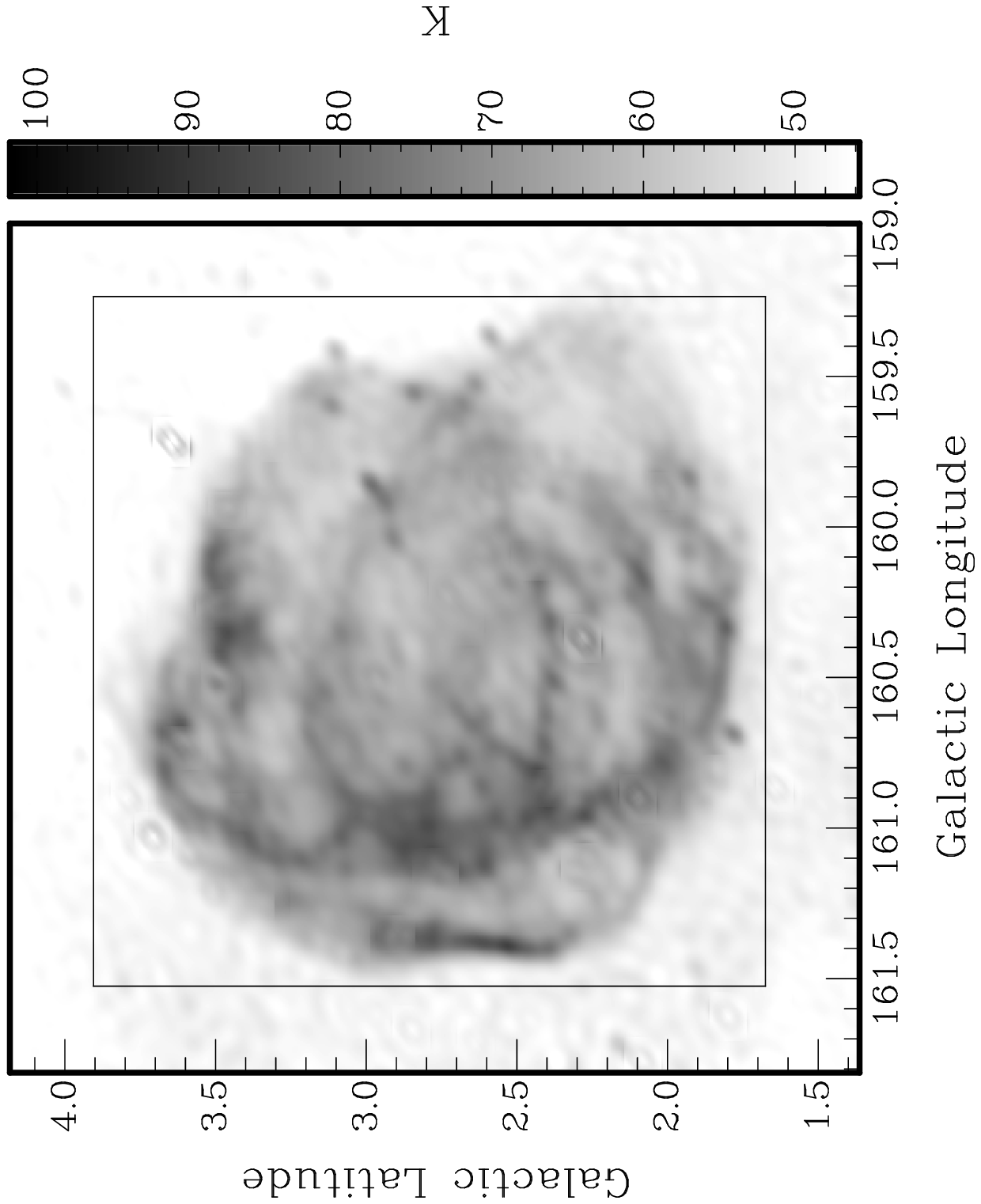}}
\put(180,260){\includegraphics{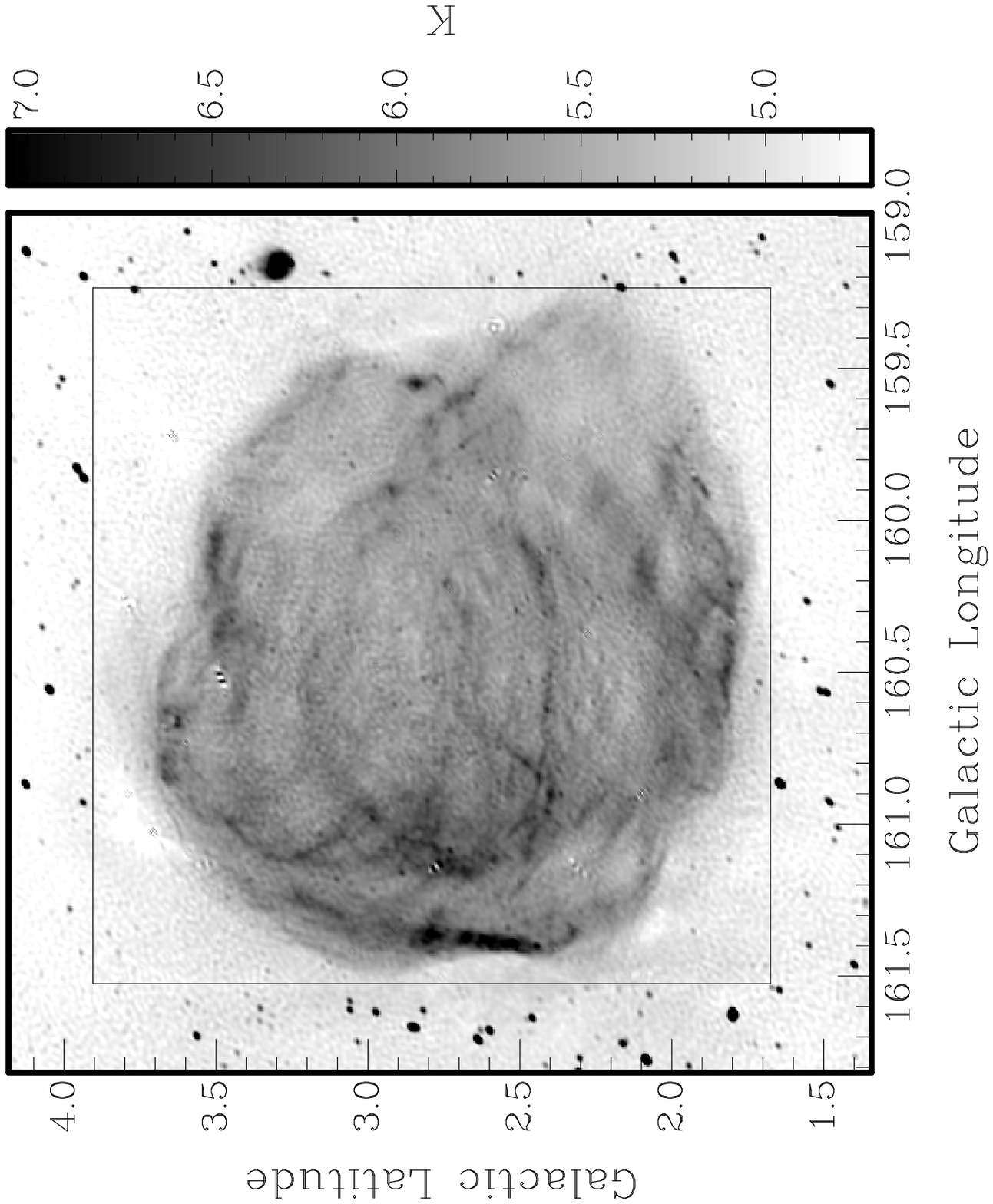}}
\end{picture}
\caption[xx]{The first row shows HB9 at
408 MHz (left) and 1420 MHz (right). The second row shows the  
408 MHz image with compact sources (CS) subtracted (left) and
the 1420 MHz image with CS subtracted (right). The single large box is the area used for 
T-T plots of the whole of HB9;
the 52 small boxes labeled with numbers show areas used for T-T plots of subareas of HB9. 
}
\end{figure*}

\begin{figure*}
\vspace{40mm}
\begin{picture}(70,100)
\put(-100,270){\includegraphics{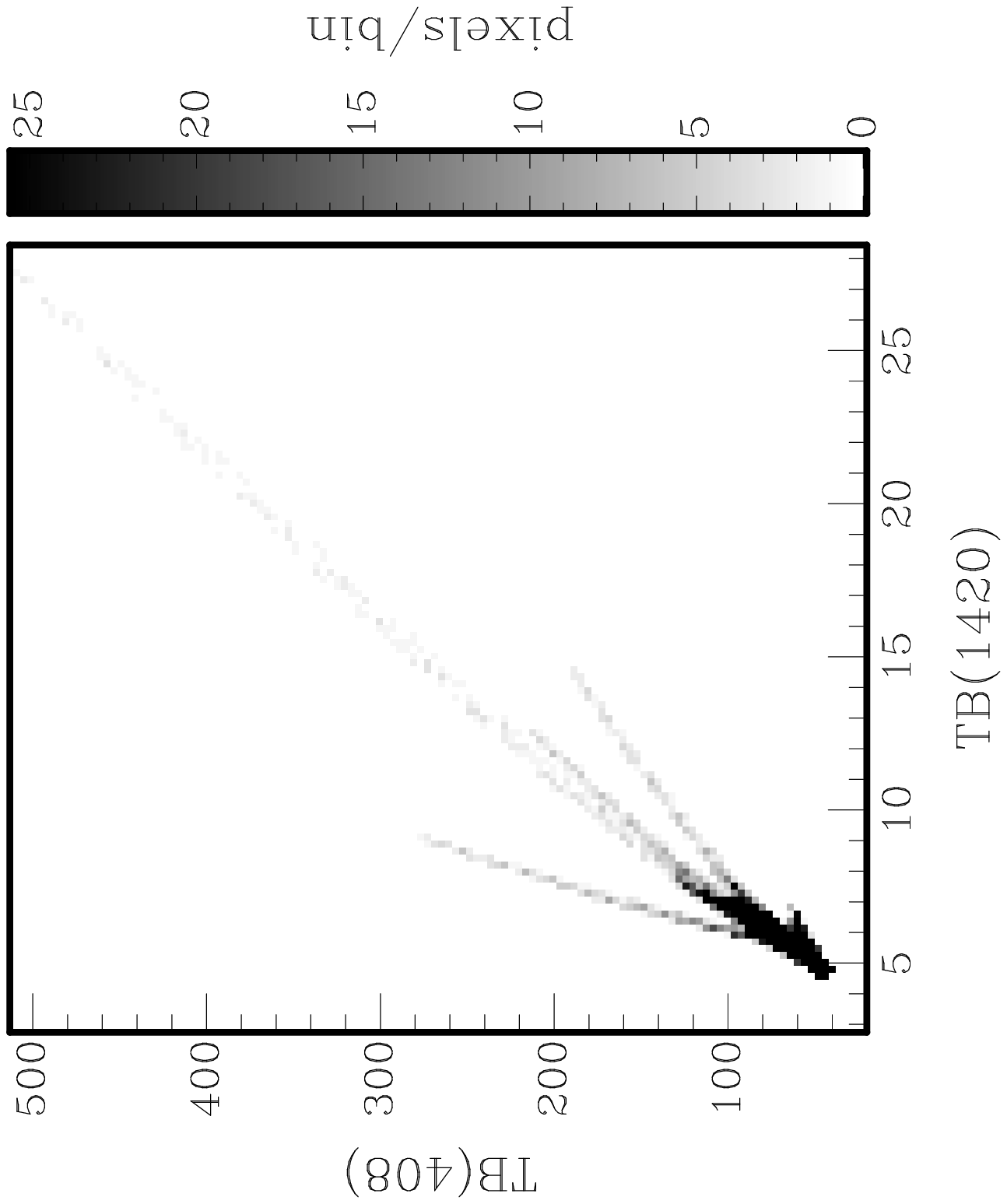}}
\put(280,218){\includegraphics{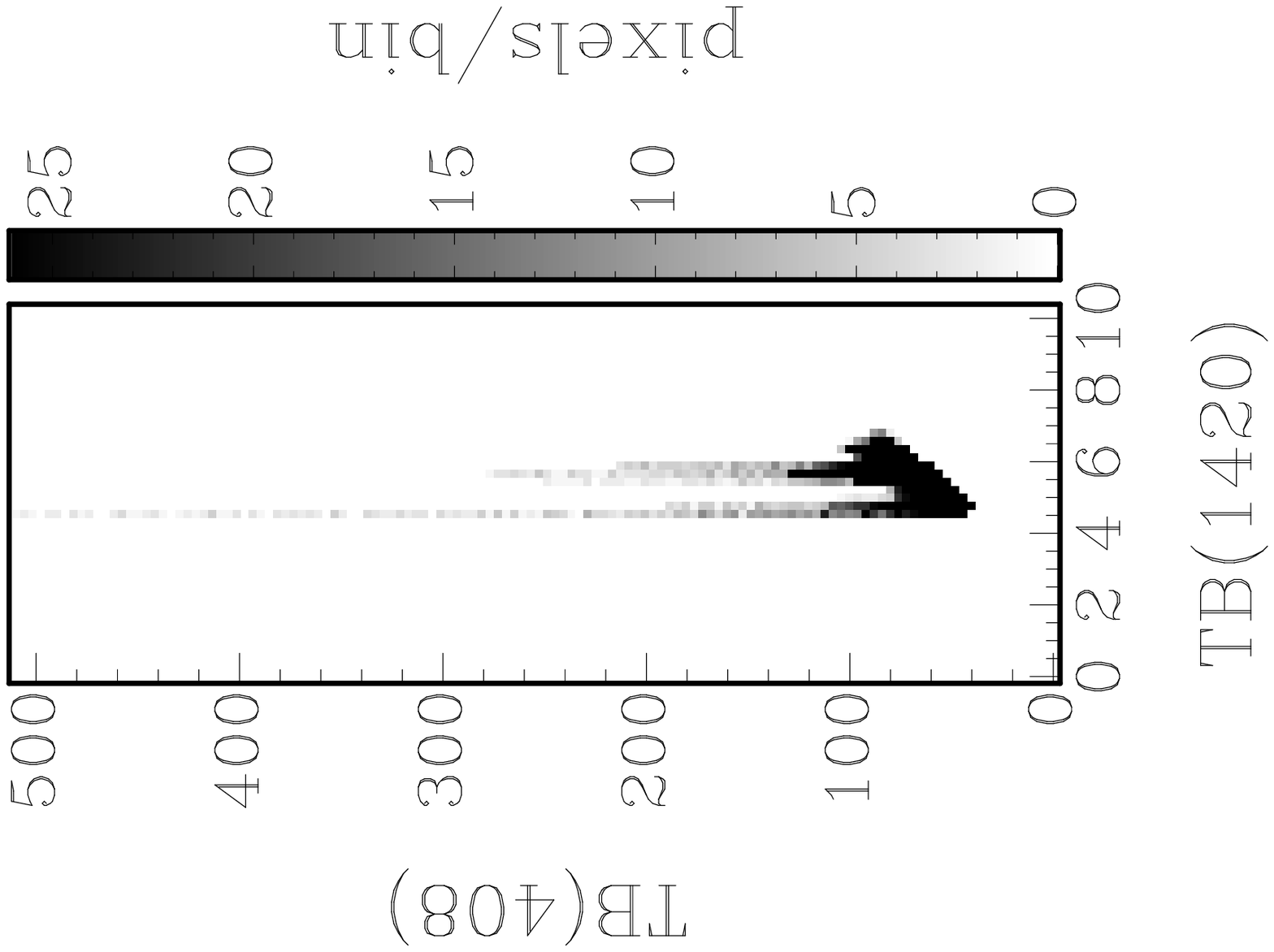}}
\put(137,212){\includegraphics{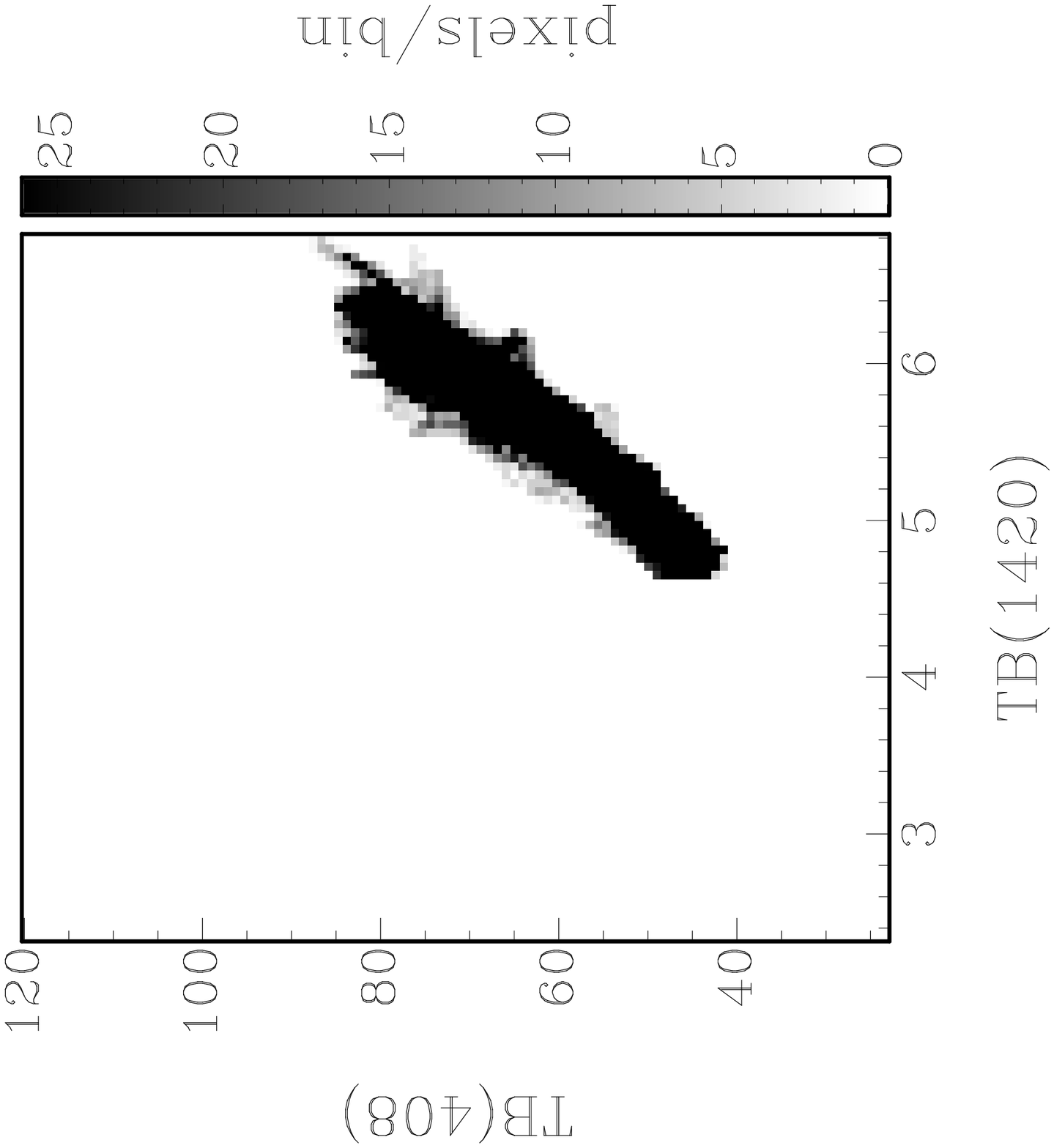}}
\end{picture}
\caption[xx]{408 MHz-1420 MHz T-T plots for the whole of HB9: 
including CS (left, $\alpha_{auto}$=0.50$\pm$0.03); 
for CS subtracted from both 408 MHz and 1420 MHz maps (middle, $\alpha_{auto}$=0.48$\pm$0.03);
for CS subtracted from 1420 MHz map only (right, $\alpha_{manual}$=0.49) }
\end{figure*}

\begin{figure*}
\vspace{30mm}
\begin{picture}(60,100)
\put(-50,185){\includegraphics{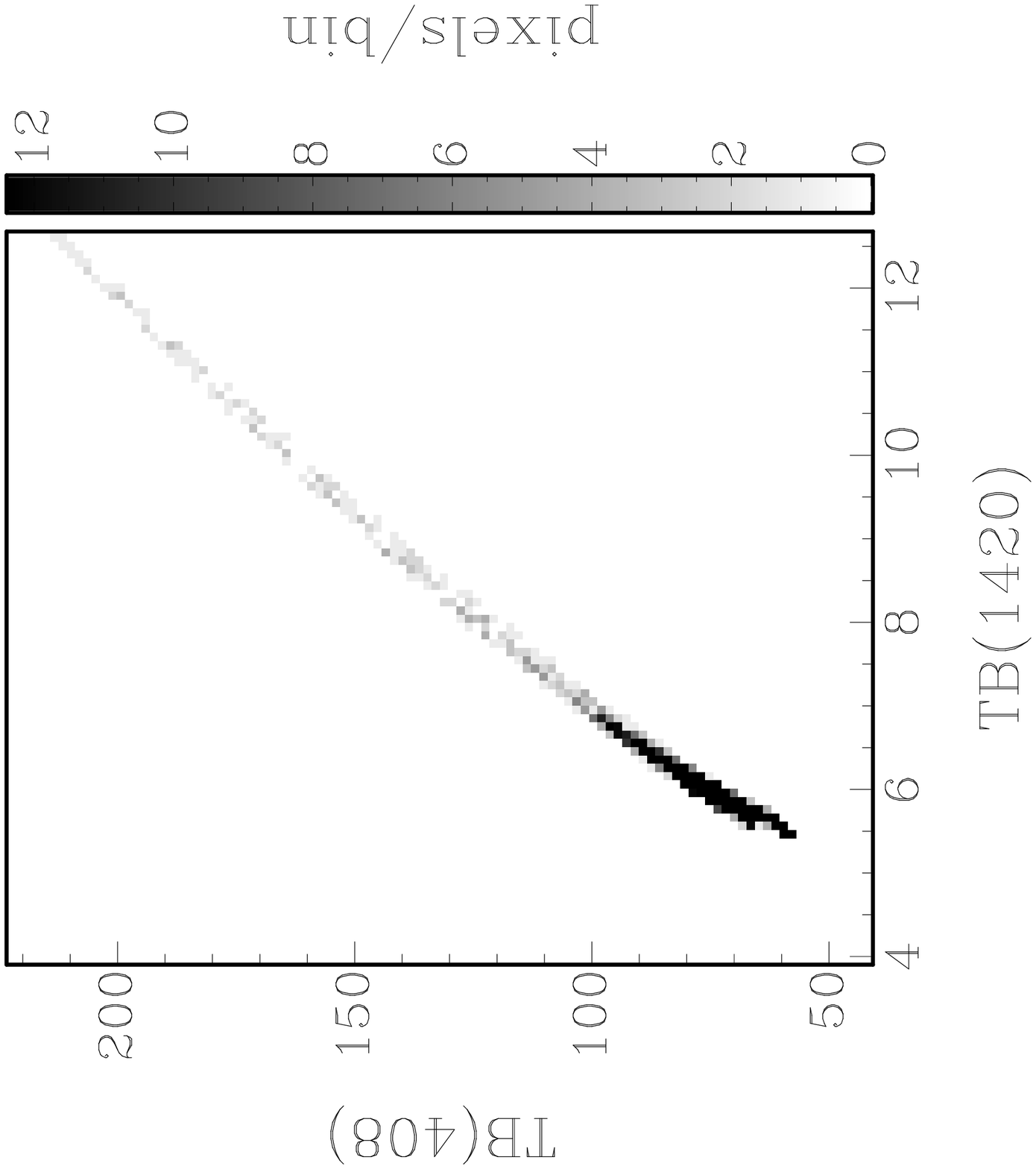}}
\put(128,188){\includegraphics{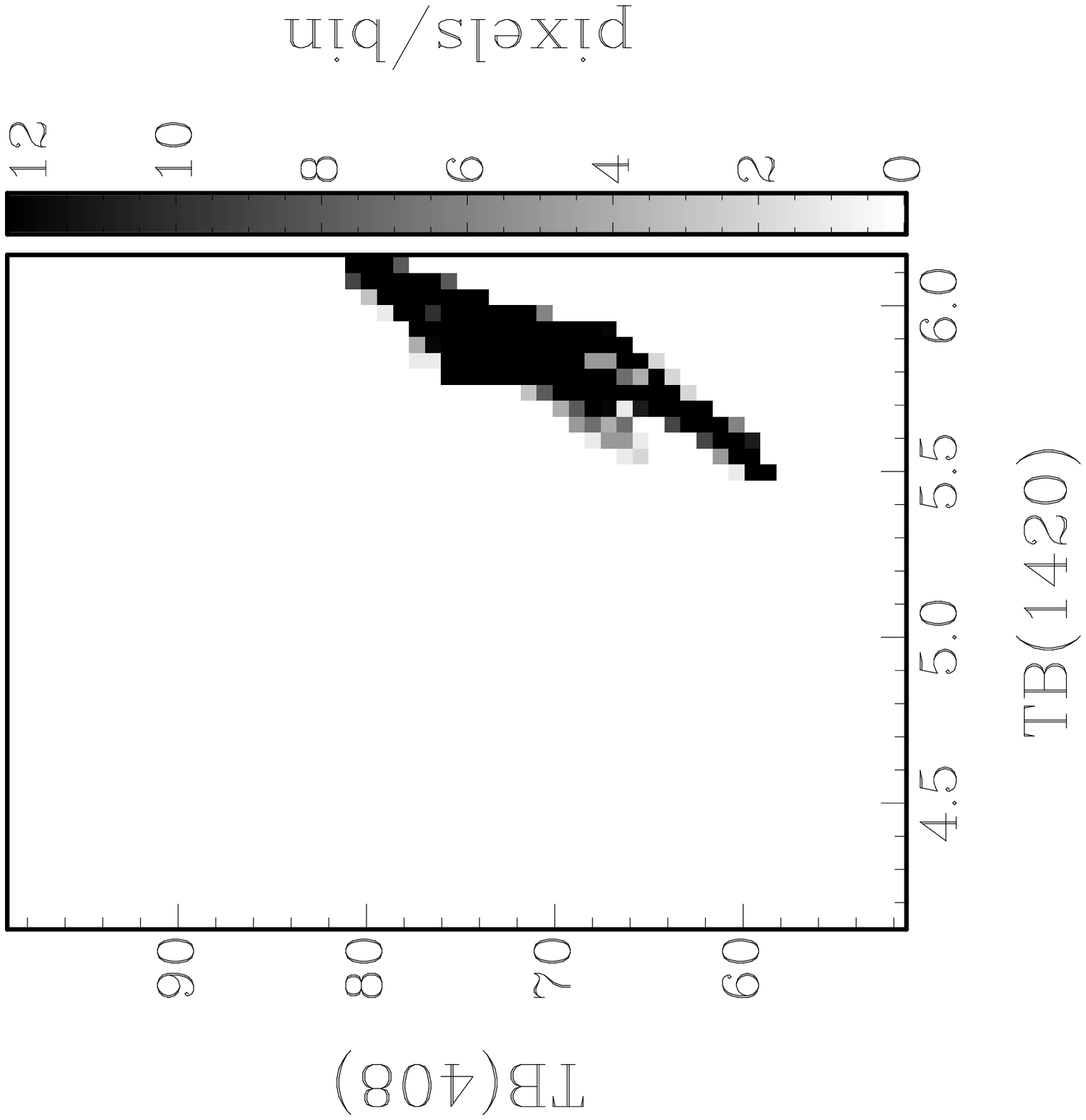}}
\put(260,180){\includegraphics{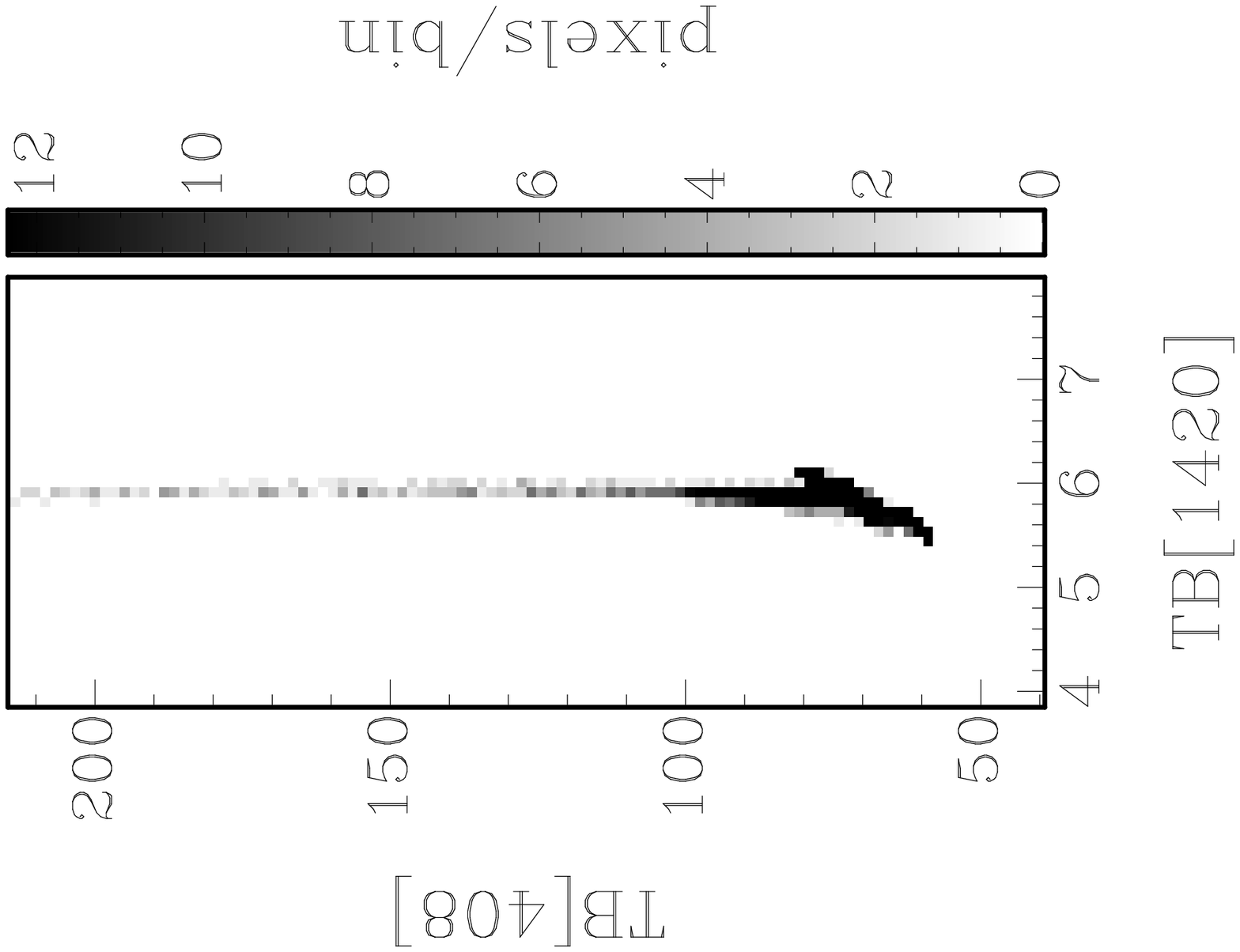}}
\end{picture}
\caption[xx]{408 MHz - 1420 MHz T-T plots of region 21: including CS (left); 
for CS subtracted from both 408 MHz and 1420 MHz maps (middle);
for CS subtracted from 1420 MHz map only (right).}
\end{figure*}

\begin{figure}
\vspace{30mm}
\begin{picture}(60,100)
\put(0,0){\includegraphics{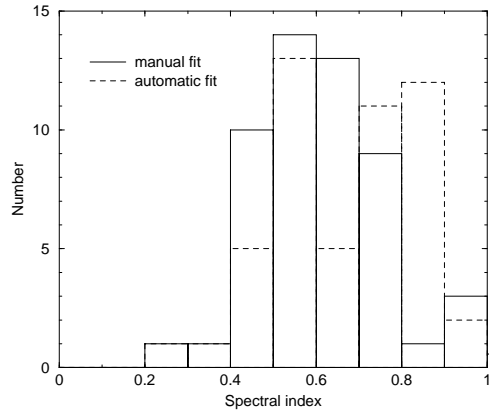}}
\end{picture}
\caption[xx]{Histogram of spectral indices in HB9: 
automatic fit includes compact sources (CS); manual fit excludes CS.}
\end{figure}

\begin{table}
\begin{center}
\caption{HB9 408-1420 MHz Spectral Index for Regions 1 to 52}
\setlength{\tabcolsep}{1mm}
\begin{tabular}{cccccc}
\hline
Region  & $\alpha_{case1}$ & $\alpha_{case3}$ &Region  & $\alpha_{case1}$ & $\alpha_{case3}$\\
\hline
 1& 0.58$\pm$0.01& 0.55 & 27& 0.75$\pm$0.07& 0.65\\ 
 2& 0.51$\pm$0.01& 0.52 & 28& 0.85$\pm$0.11& 0.76\\
 3& 0.55$\pm$0.02& 0.53 & 29& 1.05$\pm$0.08& 0.97\\
 4& 0.50$\pm$0.01& 0.47 & 30& 1.38$\pm$0.01& 0.51\\
 5& 0.44$\pm$0.04& 0.50 & 31& 0.75$\pm$0.02& 0.53\\
 6& 0.65$\pm$0.09& 0.53 & 32& 0.52$\pm$0.02& 0.40\\
 7& 0.85$\pm$0.09& 0.63 & 33& 0.82$\pm$0.03& 0.74\\
 8& 0.87$\pm$0.09& 0.55 & 34& 0.87$\pm$0.11& 0.65\\
 9& 0.86$\pm$0.03& 0.76 & 35& 0.81$\pm$0.11& 0.69\\
 10& 0.72$\pm$0.10& 0.60& 36& 0.77$\pm$0.02& 0.72\\
 11& 0.78$\pm$0.05& 0.52& 37& 0.77$\pm$0.03& 0.63\\
 12& 0.52$\pm$0.05& 0.54& 38& 0.81$\pm$0.02& 0.80\\
 13& 0.57$\pm$0.04& 0.46& 39& 0.83$\pm$0.04& 0.69\\
 14& 0.69$\pm$0.02& 0.70& 40& 0.29$\pm$0.05& 0.33\\
 15& 0.91$\pm$0.03& 0.96& 41& 0.42$\pm$0.02& 0.41\\
 16& 0.93$\pm$0.03& 0.91& 42& 0.55$\pm$0.23& 0.55\\
 17& 0.75$\pm$0.02& 0.56& 43& 0.77$\pm$0.03& 0.73\\
 18& 0.77$\pm$0.10& 0.63& 44& 0.56$\pm$0.04& 0.61\\
 19& 0.88$\pm$0.07& 0.79& 45& 0.52$\pm$0.02& 0.55\\
 20& 0.85$\pm$0.07& 0.77& 46& 0.53$\pm$0.02& 0.55\\
 21& 0.51$\pm$0.01& 0.61& 47& 0.35$\pm$0.02& 0.29\\
 22& 0.50$\pm$0.02& 0.48& 48& 0.51$\pm$0.02& 0.50\\
 23& 0.72$\pm$0.02& 0.70& 49& 0.57$\pm$0.03& 0.46\\
 24& 0.84$\pm$0.02& 0.73& 50& 0.76$\pm$0.03& 0.69\\
 25& 0.70$\pm$0.09& 0.76& 51& 0.45$\pm$0.03& 0.40\\
 26& 0.69$\pm$0.05& 0.66& 52& 0.70$\pm$0.14& 0.50\\
\hline
\hline
\end{tabular}
\end{center}
\end{table}

\subsection{HI Emission}
The CGPS data has 2 times better velocity resolution and better sensitivity than the older DRAO HI data 
(Leahy $\&$ Roger 1991). We find HI emission which is spatially associated with 
the boundary of HB9 and a deficit of emission associated with the interior in the velocity 
range -3 to -9 km/s, but not at other velocities.
This suggests that this HI is likely to be physically associated with the SNR.  
Fig. 5 shows the average of the maps of HI emission in the 8 channels from 
-3 to -9 km/s, with a contour of continuum emission at 1420 MHz to show the boundary of HB9.

\begin{figure*}
\vspace{120mm}
\begin{picture}(80,60)
\put(-10,390){\includegraphics{hb9-hi-sum.eps}}
\end{picture}
\caption[xx]{HI emission in the field centered on HB9: average of velocity channels -3 to 
-9 km$\/$s, with 1420 MHz continuum contour at 5.3 K $T_{B}$ indicating HB9. HI image is noise 
limited with rms brightness temperature $\delta$$T_{B}$$\sim$ 1 K. The synthesized beam for the 
HI line images is 58$^{\prime}$$^{\prime}$$\times$ 80$^{\prime}$$^{\prime}$.}
\end{figure*}

\section{Discussion}

\subsection{Distance of HB9, age, and possible association with pulsar 0458+46}

For the velocity range of -3 to -9 km/s, using circular galactic rotation velocity 
V$_{R}$=220 km/s and solar distance R$_{0}$=8.5kpc from the Galactic center, 
yields a distance to HB9 of d=0.8$\pm$0.4 kpc.   
This distance is consistent within errors with previous estimates: greater than 1 kpc derived from measurements 
of radial velocity of H$\alpha$ filaments by Lozinskaya (1981); about 1.1 kpc from the 
X-ray properties of HB9 by Leahy (1987); and 1.3 - 1.8 kpc from surface-brightness-diameter relations 
by Milne (1979) and Caswell $\&$ Lerche (1979).  

Since the angular size of HB9 is 130$^{\prime}$ by 120$^{\prime}$, 
the mean radius is 15 pc at 0.8 kpc.  
A Sedov model (Cox D., 1972) is applied, using the X-ray temperature and X-ray flux from
Leahy and Aschenbach (1995). For a 15 pc radius, an age of $6600$ yrs and a ratio between 
explosion energy $E_{51}$(in $10^{51}$erg) and initial density $n_0$ 
(in $cm^{-3}$) $E_{51}$/$n_0$=5 are obtained. 
 
As noted before by Leahy and Aschenbach (1995), based on the X-ray morphology and uniform
temperature, it is likely that Sedov model does not apply in
this case. The evaporative cooling model developed by White $\&$ Long (1991)
provides a better explanation for the observed X-ray features. If White $\&$
Long's (1991) model is applied for a distance of 0.8 kpc, a range of
parameters is found which give a nearly flat temperature profile at 0.8 keV and an 
appropriately centrally brightened X-ray profile
($\tau=10$ and C=20 to C=50). Then the
preshock intercloud density is 3-10$\times10^{-3}$cm$^{-3}$, 
the age is 4000-7,000 yr and the explosion energy is 0.15-0.3$\times10^{51}$erg.

The radio pulsar 0458+46 is only 23$^{\prime}$ from the center of HB9. 
It has a DM distance 1.8 kpc, spin-down age $1.8\times10^6$yr and transverse
velocity 95.5 km/s (Manchester et al., 2005).
The DM distance is expected to be larger than the distance derived from the HI data, 
due to the extra DM introduced by the extra electron density in the SNR shell.
The spindown age is based on the assumption of a fast initial spin, so it probably overestimates the true age. 
The pulsar's kinematic age based on its transverse velocity is $5.5\times10^4$yr, several
times larger than HB9's age from either Sedov or evaporative cloud models. 
Thus an association between the pulsar and HB9 is possible, but only if it was a somewhat
off-center explosion: for instance, the kinematic age can be reduced to about 7,000 yr if the explosion
center is $\sim3^\prime$ from the current pulsar position. Better X-ray observations of HB9 
should determine the nature of SNR, and give an improved SNR age value.

\subsection{HB9 Spectral Index}

\begin{figure*}
\vspace{40mm}
\begin{picture}(70,100)
\put(-10,-80){\includegraphics{specmap2.1.eps}}
\put(245,-95){\includegraphics{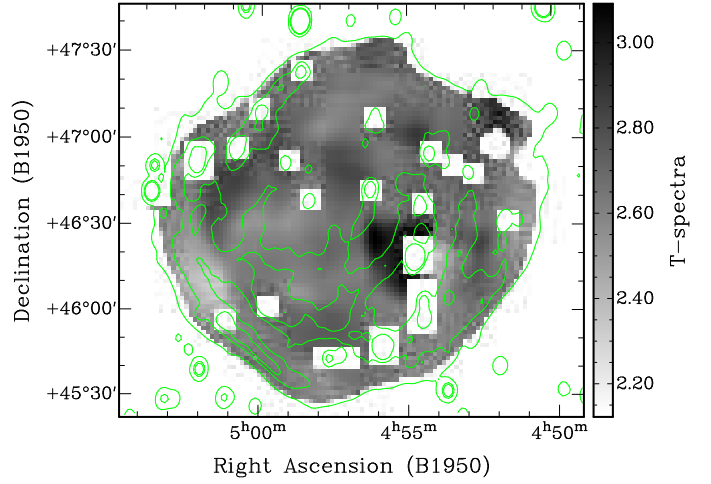}}
\end{picture}
\caption[xx]{408-1420 MHz spectral index maps of HB9: from the current images with CS
subtracted at both frequencies (left panel) with 1420 MHz 
contours at 5.25, 5.7, 6.1 K; the spectral index map from the old 408 and 1420 MHz images (right panel,
same as Fig. 6 of Leahy et al. (1998).}
\end{figure*}

Based on T-T plots, Leahy $\&$ Roger (1991) obtained a mean T-T plot spectral index of 0.61 for HB9. 
The new T-T plot spectral index derived for the whole SNR HB9  
($\alpha$=0.48$\pm$0.03) is much smaller, but consistent with the integrated 
flux density-based spectral index ($\alpha$=0.47$\pm$0.06). Two main factors appear as 
responsible for this. 
First, there is a significant change (3-4 K difference) in SNR 
brightness temperature at 1420 MHz compared to earlier data set. Fig. 2 of Leahy $\&$ Roger (1991) 
shows a bright filament at $\sim$ 3.2 K compared to $\sim$ 6.5 K in the current 1420 MHz image,  
and no such change in the 408 MHz images.
Second, the present observations have higher resolution and 
sensitivity, so more compact sources have been resolved and subtracted from the images of HB9. 
      
In what follows, we discuss the spectral index variations with location within HB9.  
Since the integrated flux and T-T plot spectral indices for the whole of HB9 agree each other, 
the filamentary emission (measured by the T-T plot method) and the total emission 
(filamentary plus spatially smooth emission) should have the same spectral index.  
To analyze local variations we divided the 52 regions into 3 groups: 
(a) boxes that cover regions with strong filamentary emission (those labeled with numbers 2, 3, 
4, 5, 7, 8, 9, 10, 11, 14, 19, 32, 33, 47, 50);(b) boxes overlapping weak filamentary emission 
(boxes 15, 16, 20, 26, 27, 30, 34, 35, 36, 37, 41, 42, 45, 51, 52); and (c) those with intermediate 
filamentary emission (the remainder). The means (and standard deviations) from the 3 
groups were 0.58 (0.14), 0.64 (0.17) and 0.61 (0.14), respectively. Thus there is no 
clear difference in spectral index between regions 
including strong filaments and regions with weak filaments.  We can therefore
conclude that the spectral results confirm the scenario suggested by Leahy $\&$ Roger (1991), 
where the filaments are edge-on viewing of emitting sheets.

Larger differences in spectrum are apparent from the comparison of the SNR 
periphery with regions in the interior. When one looks at spectral index 
($\alpha_{case3}$ in Table 1) vs. position, one finds that 
the flatter spectral indices ($<0.5$) are all associated with the SNR limb and all of the
steeper values ($>0.7$) are associated with the interior. The mean and standard deviation of spectral
indices for all limb regions are 0.49 and 0.11, whereas the values for the interior regions
are 0.67 and 0.12, respectively. An alternate analysis of spectral index was carried out to verify this result.
We made a spectral index map of HB9 using the running T-T plot method of Zhang et al. (1997):
the method yields $T_B$ spectral index $\beta$, with $\beta=\alpha+2$.
We used the CS subtracted images at 408 MHz and 1420 MHz to create the 
spectral index map shown in Fig. 6, left panel. 
The small white patches show the subregions where the fitting algorithm did not obtain a reliable value 
of spectral index. The spectral index map computed from images including
CS is similar except it has a number of elliptical spots (one for each bright CS)
at the locations of the CS and with shading corresponding to the CS spectral index.
One can see from Fig. 6 (left) that the limb of HB9 has predominantly low spectral index and the
interior has predominantly high spectral index. From the discussion here and in previous papers (e.g.
Tian $\&$ Leahy (2005), Leahy (2006)), the reliability of the automatic fits used by the spectral
map software is often poor, yet the map can give a general indication of spatial variations of
spectral index.
 
A general discussion of mechanisms that govern the radio spectrum is given in Leahy \& Roger (1998).
Absorption processes and electron energy losses from ionization act more effectively at lower 
frequencies and result in spectral flattening. Synchrotron and inverse-Compton act at high energies
and steepen the spectrum. If the electron energy spectrum is curved
then spatial variations in spectral index can occur due to variations in magnetic field,
which determines the observed synchrotron frequency for a given electron energy.
The diffuse galactic electron spectrum has an electron energy index steepening from $\sim$2.0 
below $\sim2$GeV to $\sim$2.6 above $\sim5$GeV, then to $\sim$3.5 above $\sim40$GeV, corresponding
to synchrotron index increasing from $\alpha$=0.5 at low frequencies to $\alpha$=1.25 at high frequencies. 
For a fixed observing frequency, increasing magnetic field, B, means sampling lower electron energies,
so that spectral index flattens with increasing B.

For HB9, the steeper index for interior regions, for which the line of sight goes through the whole SNR,
could be interpreted as due to lower B for the interior compared to the rim; stronger synchrotron losses
for the interior; or stronger absorption effects at the rim. For the Cygnus Loop (Leahy \& Roger, 1998)
low frequency observations confirm that the NE rim shows the effects of thermal plasma absorption. However
for HB9, previous studies have not found significant weakening of the radio brightness at the rim at
low frequencies. This leaves the first two causes to examine more closely. The radio morphology of HB9 is 
consistent with a low brightness interior surrounded by a brighter shell with filaments: the image at
1420 MHz in Figure 1 show this most clearly. The highest compression in an SNR should occur just inside
the forward shock, where the optical filaments form. Thus one expects the highest magnetic fields
and highest synchrotron emissivity in the high compression regions. This now explains the 
observed spectral index variations: higher magnetic field at larger radius within the SNR produces flatter
radio spectrum for the outer regions. Thus a larger proportion of steep spectrum emission for 
lines-of-sight through the central
body of the SNR results in an observed steeper spectral index for interior regions.  
The conclusion is consistent with the explanation given by Leahy et al. (1998) for spectral index 
variations
at HB9, although that study did not do a comparison of limb and interior spectral index values.
 
The study of Leahy et al. (1998) used the old DRAO 408 MHz
and 1420 MHz observations and combined those with existing 2695 MHz and 4850 MHz data from the Effelsberg
telescope, observations at 151 MHz with the Cambridge Low-Frequency synthesis telescope and 232 
MHz observations with the Beijing Astronomical Observatory Miyun telescope. 
The 151 and 232 MHz maps suffered from lack of low-order spacings, and
the mean derived spectral indices for 151-232 MHz and 232-408 MHz indicate that the brightness scale for
232 MHz map was too large. 
To compare the effect on spectral index of the improved 408 and 1420 MHz data, we show the old 408-1420
MHz spectral index map in Fig. 6 (right panel) beside the spectral index map derived from the new data.
The old map has an important artifact produced by the point source: the large dark area near  
RA 4H55M, Dec +46d20m is due to 4C46.09. 
The contribution removed in that analysis was apparently too small to take into account the high brightness of 
4C46.09. The same artifact affects the spectral index map made using the new data with CS included, 
but not in the spectral index map (shown in the left panel of Fig. 6) with CS subtracted.
The old and new spectral index maps are globally similar (e.g. the limb has a flatter spectral 
index than the interior, with the lowest index coming from the bright southeast filament).  
There are, however, significant differences in the details between the two spectral index maps.
This might be attributed to the sensitivity of the automatic fitting routine to small changes: for example
compare case 1 (automatic fit) and case 3 (manual fit) in Table 1.
It is interesting to note that the frequency-averaged spectral index map, which should suffer least from 
these errors, bears similarity to the results presented here: steeper index from the central regions than the rim.  
 
\section{Conclusion}
We present new higher sensitivity and higher resolution images of the SNR HB9 at 408 MHz and 1420 MHz, 
and study its radio spectrum, corrected for compact source flux densities using new improved methods. 
The T-T plot spectral index for HB9 agrees with the integrated flux-density based 408-1420 MHz spectral index. 
A study of spatial variations shows no systematic difference in spectral index between weak and strong filaments, 
supporting the conclusion of Leahy $\&$ Roger (1991). 
We find steeper spectral index for interior regions than for rim regions. 
This can be explained by a standard curved interstellar electron energy spectrum combined with
variable magnetic field. Due to lower compression, a lower magnetic field in the interior,
compared to that near the outer shock, results in steeper spectrum emission from the interior.
Thus a larger proportion of steep spectrum emission for lines-of-sight through the central
body of the SNR results in an observed steeper spectral index for interior regions.
Based on HI features associated with HB9, we obtain a distance of 
0.8$\pm$0.4 kpc, and give updated Sedov and evaporating cloud model parameters for the SNR.
We discuss the possible association between the pulsar and HB9, and conclude 
that more evidence is necessary to decide this possible association.
 
\begin{acknowledgements}
We acknowledge support from the Natural Sciences and Engineering Research Council of Canada. WWT thanks the NSFC for support. 
The DRAO is operated as a national facility by the National Research Council of 
Canada. The Canadian Galactic Plane Survey is a Canadian project with international partners. 
\end{acknowledgements}

\end{document}